**In-Person:** Oral ☐ / Poster ☐ / The same ☐

**Virtual:** Zoom ☐ / Pre-recorded video ☒

**Topic:** *Unmanned Aircraft Vehicles (UAV) and Systems*

# A Tilting-Rotor Enhanced Quadcopter Fault-Tolerant Control Based on Non-Linear Model Predictive Control

Yanchao Wang [1], Xu You [2] and Mehdi Baghdadi [1]

[1] University College London, Advanced Propulsion Lab, London, United Kingdom
[2] Wuhan University of Technology, State Key Laboratory of Maritime Technology and Safety, Wuhan, China
E-mail: yanchao.wang.20@ucl.ac.uk

**Summary:** This paper proposes a fault-tolerant control strategy based on a tilt-rotor quadcopter prototype, utilizing nonlinear model predictive control to maintain both attitude and position stability in the event of rotor failure. The control strategy employs an extended state observer to predict model deviations following a fault and adjusts the original model in the subsequent time step, thereby achieving active fault-tolerant control. The proposed method is evaluated through simulations and compared to both traditional quadcopter and tilt-rotor quadcopter without observer under identical conditions. The results demonstrate that the tilt-rotor quadcopter can maintain position control without sacrificing yaw stability, unlike traditional quadcopters.

## 1. Introduction

A quadcopter is a multi-rotor aircraft powered by four sets of actuators, which include motors, electronic speed controllers (ESCs), and propellers. With advancements in control algorithms, quadcopters have found extensive applications across various fields. However, without redundancy, actuator failures of quadcopters present a significant challenge. Such failures can compromise the robustness of the quadcopter or even result in uncertain damage. As a result, fault-tolerant control (FTC) of quadcopters has garnered significant attention from both the industry and academia, and has been extensively discussed in several previous programs [1] – [7].

The most widely adopted controllers in Fault-Tolerant Control (FTC) strategies are Nonlinear Model Predictive Control (NMPC), Prescribed Performance Control (PPC), and Sliding-Mode Control (SMC), each designed based on the controlled object model. NMPC, despite typically having slower computational performance compared to SMC, offers significant advantages by reducing the complexity associated with linearization and attitude decoupling of the controlled object. This makes NMPC a highly versatile approach, particularly in scenarios where precise trajectory tracking and dynamic adaptability are required. Furthermore, optimization-based framework of NMPC allows it to account for input constraints and system uncertainties, leading to enhanced performance and robustness in applications. FTC methodologies are generally classified into two categories: passive and active. Active FTC dynamically adjusts controller parameters or updates the system model in real time based on fault detection and diagnosis. This is achieved by observers, redundant sensors, or adaptive algorithms that continuously monitor the system's performance and compensate for detected faults. By integrating active FTC strategies, modern autonomous systems can maintain stability and performance even with actuator failures, sensor malfunctions, or unexpected disturbances. The tilt-rotor quadcopter (TRQ) is a class of deformable drones characterized by actuator groups that can be independently tilted through motor servo systems. Compared to standard quadcopters, TRQs offer additional control inputs, effectively addressing the under-actuation and strong coupling challenges commonly associated with conventional quadcopters. The incorporation of redundant controllers enhances their anti-interference capabilities and provides passive fault tolerance, thereby improving overall robustness and reliability.

The program proposed in this paper builds upon prior prototypes [5] and is inspired by the experimental work of Santos et al. [3] and Jiali et al. [4]. Santos employed the Unscented Kalman Filter (UKF) to estimate the mathematical model of the drone within the framework of NMPC, enabling a tilt-rotor dualcopter (TRD) to perform trajectory tracking under the influence of an unknown hanging payload. This scenario introduced significant control challenges due to the dynamic shift in the center of gravity. However, the project simplified the modeling of the TRD nut without evaluating if the actual actuator parameters were sufficient to meet allocation requirements. Jiali et al. utilized state feedback through an Extended State Observer (ESO) to compensate for uncertainties in the mathematical model of the drone within NMPC, successfully tracking the attitude of a quadcopter equipped with three lift actuators and a single torque actuator. While the above approach demonstrated effective control when a lift actuator failed, it did not account for the positional dynamics of the quadcopter, the coupling effects between angular changes and positional shifts, or active fault tolerance in cases of torque actuator failures. This study builds upon the stability analysis of quadcopter ESO conducted by Tong et al. [6], leveraging these insights to develop a more robust fault-tolerant control strategy.

## 2. Implementation

The TRQ is a multi-input and multi-output (MIMO) system with eight inputs and twelve states. Previous works [5] shows in Fig. 1 developed a TRQ prototype incorporating a high-level controller running NMPC for control allocation, complemented by a low-level controller responsible for precise motor group coordination and execution.

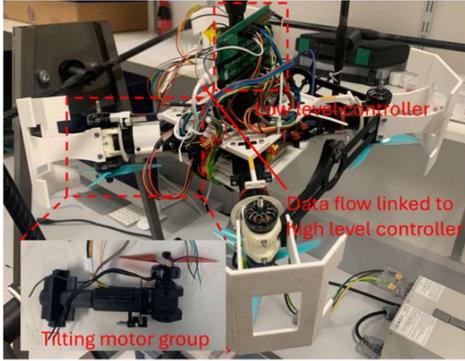

**Fig. 1** The TRQ prototype from the previous works

The inputs $u$ and the system states $x$ of TRQ can be defined as:
$$u = [\alpha^T \quad \zeta^T]^T \epsilon \mathbb{R}^8$$
$$x = [p_W^T \quad v_W^T \quad \eta_W^T \quad \dot{\eta}_W^T]^T \epsilon \mathbb{R}^{12} \quad (1)$$

Where $\zeta$ denotes the throttle percentage and $\alpha$ denotes the tilting angle of the motor groups. $p_W \epsilon \mathbb{R}^3$ denote the position in the word frame, and the velocity is described as $v_W \epsilon \mathbb{R}^3$. Based on the body frame, $\omega_B$ describes the angular velocity, and the attitude angle is denoted by $\eta_B$. All the frames are illustrated in Fig. 2.

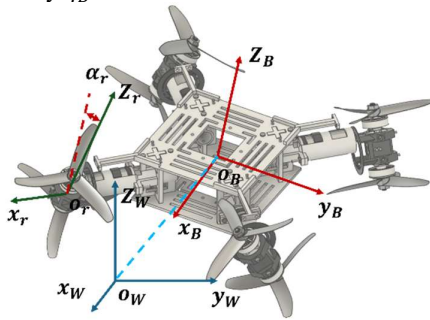

**Fig. 2** The TRQ coordinates in TRQ modelling

The expressions between throttle percentage and propeller thrust and torque are modeled as:
$$F_r = 0.0007\zeta^2 + 0.0157\zeta - 0.1891$$
$$\tau_r = 2e - 6\zeta^2 + 0.0007\zeta - 0.008 \quad (2)$$

The TRQ dynamics are then modeled as:
$$\begin{bmatrix} m & 0 \\ 0 & I \end{bmatrix} \begin{bmatrix} \dot{v}_W \\ \dot{\omega}_B \end{bmatrix} = \begin{bmatrix} F_W \\ \tau_B \end{bmatrix} - \begin{bmatrix} \omega_B \times mv_W \\ \omega_B \times I\omega_B \end{bmatrix} + \begin{bmatrix} mg \\ 0 \end{bmatrix} + d \quad (3)$$

Where $m = 2.1$ is the diagonal mass matrix of the quadcopter and $g$ is the gravity scalar. $I$ is a diagonal matrix of the moment of inertia considering the TRQ to be symmetrical, where $I_{xx} = I_{yy} = 0.01241$ and $I_{yy} = 0.02365$. $l = 0.23$ is the distance between the propeller and the body center. $F_W$ is the force mapping on the world frame. While $F_B$ and $\tau_B$ are the propeller force $F_r$ and torque $\tau_r$ mapping on the body frame. from the motor frame by a frame rotation, which is affected by the tilting angle $\alpha_{ri}$.

Above all, the nonlinear discrete state space equation can be denoted as:
$$\dot{x}_{k+1} = f(x_k, u_k, d_k)$$
$$y_{k+1} = h(x_k) \quad (4)$$

Where $y_{k+1} \epsilon \mathbb{R}^{12}$ denoted the output of the system, representing the position and angular information of the TRQ, $d_k$ represented the disturbance.

## 3. Methodology

For the simulation, only the high-level controller responsible for solving the NMPC problem is active. The NMPC controller receives the ESO-modified states relative to the reference as input and outputs the throttle percentage and motor group tilting angle. By estimating an extended state of the system based on the NMPC output and model response, the ESO generates compensatory values to reduce system error. However, in the experimental test, the ESO-estimated values will be fused with sensor feedback for improved accuracy. Fig. **3** illustrate the control flow of the system, where NMPC controller will run on a high level controller and send the output commend to the low level motor group controller.

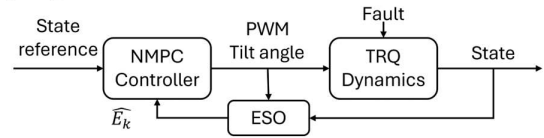

Fig. 3 Structure diagram of FTC

The MPC is performed by minimizing the cost function containing constraints to find finite-time optimal control. The cost function to minimize the control effort is given below:
$$J(x_t, u_t) = \sum_{t=k}^{k+N-1} [(x_t - x_t^r)^T Q (x_t - x_t^r) + u_t^T R u_t] \quad (5)$$

Where $k$ is defined to represent each prediction step. The reference state is denoted by $x_t^r$, and the prediction horizon is denoted by $N$. $Q$ and $R$ are the weight matrix relative to state and control.

The constrained optimal control problem in the prediction horizon is sampled in $N$ steps. At each time $t$, the problem can be denoted as:
$$\min_{x_0 \ldots x_N, u_0 \ldots u_{N-1}} J \quad (6)$$

subject to:
$$\dot{x}_{k+1} = f(x_k, u_k, e_k)$$
$$x_k \in \mathbb{X}$$
$$u_k \in \mathbb{U} \quad (7)$$

Where $k = 0,1,\ldots,N-1$. $e$ is the compensation from ESO, $x$ and $u$ are the states and inputs constraints which denoted as:

$$\mathbb{X} = \{x_{min} \leq x_k \leq x_{max}\} \\ \mathbb{U} = \{u_{min} \leq u_k \leq u_{max}\} \quad (8)$$

The NMPC method will generate a sequence of inputs u as $\{uk|k, uk+1|k, \ldots, uk+N-1|k\}$, and the first value $\{uk|k\}$ will be taken as the system inputs at every time step.

### 3.1. Control Allocation

The control allocation problem deals with finding the appropriate actuation motor speeds and tilting motor angles suitable for Eq. 1. By rearranging Eq. 2 and 3, the direct inversion control allocation can be obtained:

$$u = G^{-1}(F_W, \tau_B) \quad (9)$$

In this paper, CasADi solver is used to solve this quadratic programming problem.

### 3.2. Extend State Observer

In this work, extended states observer is introduced to estimate actuator faults, expressed as:

$$\begin{aligned}\dot{\hat{x}}_1 &= \hat{x}_2 + \beta_1 g_1(e) \\ \dot{\hat{x}}_2 &= \hat{x}_3 + \beta_2 g_2(e) + b_0(u) \\ \dot{\hat{x}}_3 &= \beta_3 g_3(e) \\ \hat{y}_k &= C\hat{x}_k\end{aligned} \quad (10)$$

Where $\hat{y}_k$ is the estimated output when the disturbance and actuator fault is considered. $b_0$ is the estimation of $b$. $\hat{E}_k = \hat{y} - \hat{y}_k$ represent the disturbance and actuator fault. Function g is taken as:

$$g_i(e) = |e|^{\alpha_i} sgn(e)$$

Where $0 < \alpha_i < 1$ is the weight of estimation.

During the process, the ESO is integrated with the TRQ model to form a modified model, expressed as:

$$\begin{aligned}\dot{\hat{x}}_k &= f(x_k, u_k, d_k) + b(u_k) \\ \hat{y}_k &= h(\hat{x}_k)\end{aligned} \quad (11)$$

Where b is denoting the uncertainty in modelling and $\hat{y}_k$ is the estimated output when the actuator fault is considered. Based on the project from Ke et al. [12], the error between the estimated and actual output will also be considered into the cost function as:

$$(\hat{y}_k - y_k)^T I(\hat{y}_k - y_k)$$

The optimization problem is solved at each iteration by NMPC, yielding the estimated actuator input for the next step.

## 4. Simulations

This section presents a series of experiments conducted to evaluate and compare the performance of quadcopters utilizing NMPC and ESO-based NMPC with the TRQ-based platform. The simulation includes a horizontal wind disturbance of $8\ m/s$, generating a disturbance around $2\ N$. To ensure consistency, the experimental setup follows the same configuration and trajectory parameters as outlined in previous studies [5]. A fault scenario is introduced, in which the top-right motor suffers a fifty percent reduction in speed during trajectory tracking, simulating a realistic actuator failure. Throughout the simulation, with a prediction period of $0.1\ s$, the traditional quadcopter struggles to maintain stability, ultimately leading to simulation crashes due to the inability to compensate for the motor failure. In contrast, the TRQ-based platform demonstrates improved resilience.

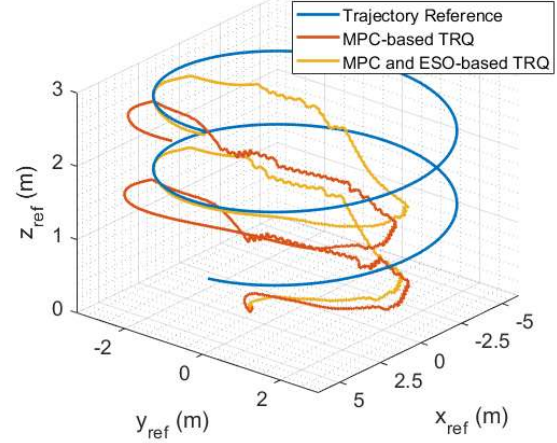

Fig. 4 The display of trajectory tracking shows the MPC and ESO-based TRQ has a better performance on fault-tolerent control.

As illustrated in Fig. 4 and Fig. 6, the ESO-based NMPC significantly enhances trajectory tracking performance under actuator faults when compared to the TRQ without an observer. These results highlight the effectiveness of ESO-based NMPC in improving fault tolerance and overall system robustness, making it a promising approach for enhancing the reliability of tilt-rotor quadrotor systems in challenging operational conditions.

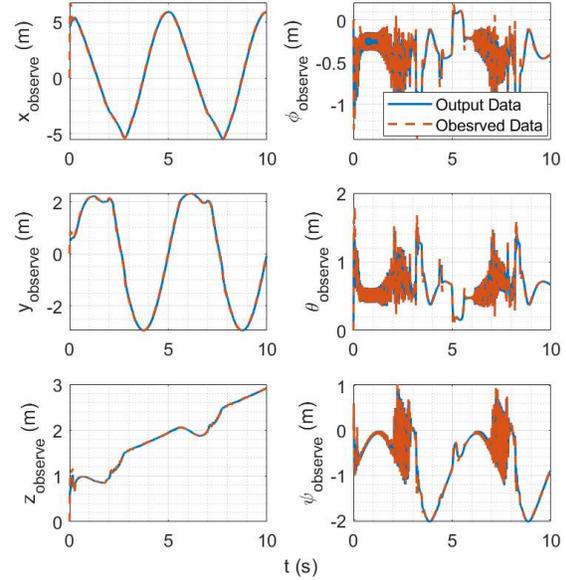

Fig. 5 The comparation between the model output and estimated ESO output

Fig. 5 illustrates the error between the model output and the ESO output, which is incorporated into the cost function of the NMPC. The results demonstrate that ESO exhibits strong tracking capabilities, effectively minimizing the deviation between estimated and actual system states. However, the ESO shows a limitation in predicting future outputs, which may impact the overall performance of the NMPC in time varying disturbance forecasting.

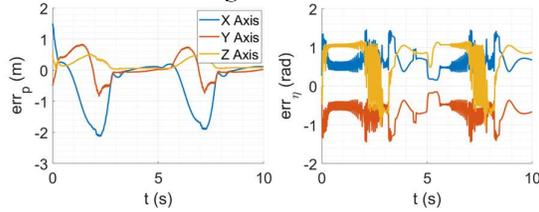

Fig. 6 The position and angle error during trajectory tracking.

To further evaluate the performance of the ESO-based NMPC, a time-varying fault is introduced to the same motor as follows:

$$\zeta_1 = \zeta_1 \times (0.8 - 0.2 \times \sin(0.01t))$$

The figure shows that with a low-frequency fault variation, NMPC can maintain TRQ stability.

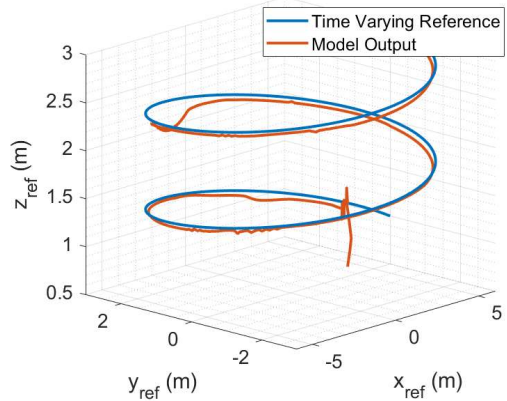

Fig. 7 The display of TRQ tracking trajectory with a time varying fault on one motor

## 5. Conclusions

This paper proposes a fault-tolerant control strategy for tilt-rotor quadrotors (TRQs) utilizing an Extended State Observer (ESO)-based Nonlinear Model Predictive Control (NMPC) framework. Building upon a previously developed prototype with a tilt-propeller mechanism, the quadrotor is modelled as a fully actuated system through the tilt actuator group method, significantly enhancing its disturbance rejection and fault tolerance. The system dynamics are derived from the actual TRQ prototype, with the NMPC inverse solution employed to compute control inputs based on reference trajectories. To mitigate disturbances, ESO is utilized for real-time state observation, compensation, and model correction. Simulation results demonstrate that the integration of TRQ and ESO enhances fault tolerance and computational efficiency, indicating its potential applicability to later prototype implementations. The current observer demonstrates better effectiveness only under disturbances and faults without large variations. Future research will integrate the Extended State Observer (ESO) with the Kalman filter to enhance its predictive capability for time-varying uncertainties within the model. Furthermore, the implementation of NMPC on a Linux platform will be explored to improve computational efficiency and real-time performance.